\documentclass[11pt,a4paper,conference]{IEEEtran}

\usepackage{graphicx}
\usepackage{amssymb}
\usepackage{caption}
\usepackage{subcaption}
\usepackage{url}
\usepackage{amsmath}
\usepackage{float}
\usepackage{caption}
\usepackage{color}

\graphicspath{{./Images/}}

\begin{document}

%
%

\title{Under the sea: Pulsing corals in ambient flow}

%
%

\author{\IEEEauthorblockN{Nicholas A. Battista\IEEEauthorrefmark{1},
Julia E. Samson\IEEEauthorrefmark{2},
Shilpa Khatri\IEEEauthorrefmark{3}, and
Laura A. Miller\IEEEauthorrefmark{4}
}
\IEEEauthorblockA{\IEEEauthorrefmark{1}Dept. of Mathematics and Statistics\\
The College of New Jersey, Ewing, NJ, USA\\
battistn@tcnj.edu}
\IEEEauthorblockA{\IEEEauthorrefmark{2}Dept. of Biology\\
University of North Carolina at Chapel Hill, Chapel Hill, NC, USA\\
jesamson@live.unc.edu}
\IEEEauthorblockA{\IEEEauthorrefmark{3}Applied Mathematics Unit, School of Natural Sciences\\
University of California Merced, Merced, CA, USA\\
skhatri3@ucmerced.edu}
\IEEEauthorblockA{\IEEEauthorrefmark{4} Dept. of Biology, Dept. of Mathematics\\
University of North Carolina at Chapel Hill, Chapel Hill, NC, USA\\
lam9@unc.edu}
}

\maketitle

%
%

\begin{abstract}
While many organisms filter feed and exchange heat or nutrients in flow, few benthic organisms also actively pulse to enhance feeding and exchange. One example is the pulsing soft coral (\textit{Heteroxenia fuscescens}). Pulsing corals live in colonies, where each polyp actively pulses through contraction and relaxation of their tentacles. The pulses are typically out of phase and without a clear pattern. These corals live in lagoons and bays found in the Red Sea and Indian Ocean where they at times experience strong ambient flows. In this paper, $3D$ fluid-structure interaction simulations are used to quantify the effects of ambient flow on the exchange currents produced by the active contraction of pulsing corals. We find a complex interaction between the flows produced by the coral and the background flow. The dynamics can either enhance or reduce the upward jet generated in a quiescent medium. The pulsing behavior also slows the average horizontal flow near the polyp when there is a strong background flow. The dynamics of these flows have implications for particle capture and nutrient exchange. 

\end{abstract}


%
%

\section{Introduction}


Some benthic organisms actively pulse to enhance particle capture and nutrient exchange by generating flow currents, such as the upside down jellyfish (\textit{Cassiopeia}) and pulsing corals (\textit{Xenia} and \textit{Heteroxenia}). Not only do these organisms actively pulse for the benefit of feeding, but pulsing has been shown to increase photosynthetic rates \cite{Kremien:2013}. Both upside down jellyfish and pulsing corals host zooxanthellae in their tissues, which photosynthesize \cite{Kaplan:1988,Fitt:1998,Welsh:2009}. Since both pulsing corals and upside down jellyfish reside in marine environments, the currents they produce are subject to perturbations from the ambient fluid motion in their environment. Previous and current studies have quantified the current produced by both the upside down jellyfish and pulsing coral in quiescent conditions \cite{Hamlet:2012,Hamlet:2014,Samson:coral,Samson:coral-Re}, as well as the upside down jellyfish with background flow \cite{Hamlet:2012b}. The addition of ambient flow was shown to significantly affect the currents produced by the upside down jellyfish \cite{Hamlet:2012b}, suggesting that the bell and oral arm morphologies are advantageous for particle capture and feeding under a range of conditions. 

\begin{figure}
\centering
\includegraphics[width=0.45\textwidth]{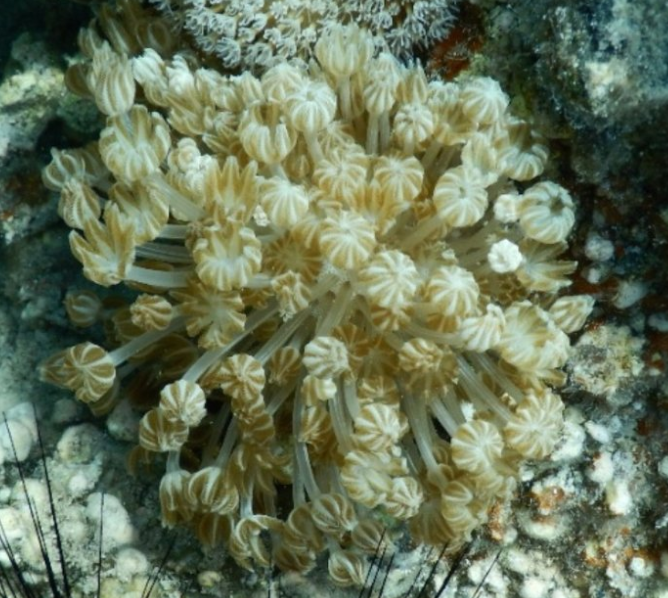}
\caption{Xenia colony in Eilat, Israel.}
\label{fig:Xenia_Colony2}
\end{figure}

There are some intriguing differences between upside down jellyfish and pulsing soft corals which make them independently worthy of study. While upside down jellyfish may remain isolated or form small groups, pulsing corals are typically found in large, dense colonies that can grow up to $60\ cm$ across, where each individual polyp is roughly $5\ cm$ tall \cite{Lieske:2004}, see Figure \ref{fig:Xenia_Colony2}. There are also morphological differences in how the flows are generated. Each coral polyp has 8 tentacles that actively contract and relax to generate flow. On the other hand, upside down jellyfish generate feeding currents by actively contracting a bell that pushes flow through an array of elaborate oral arms. Upside down jellyfish are typically found in protected and relatively stagnant areas, while soft corals are often found relatively exposed to strong flows. There are also some similarities: individual upside down jellyfish and pulsing corals both generate continuous upwards jets and do not pulse in synchrony.

In this paper, we use the immersed boundary method to simulate a three-dimensional model of a pulsing coral in a viscous fluid. Shear flow is prescribed as the background flow, and it interacts with the flow currents actively generated by the polyp. We quantify the upward jet generated by the coral over a range of background flow speeds. We also describe the average flow moving towards the coral along the substrate. 


%
%

\section{Methods}
\label{sec:methods}

To solve the fully coupled fluid-structure interaction problem of a pulsating soft coral with background flow in an incompressible, viscous fluid, the immersed boundary method (IB) \cite{Peskin:2002} was implemented. IB has been successfully applied to a variety of applications in biological fluid dynamics including flow within embryonic hearts \cite{Baird:2015,Battista:2015b}, flow past leaves \cite{Miller:2012,Zhu:2011}, swimming and flight  \cite{Miller:2009,Hershlag:2011,Hamlet:2015}, and blood clotting through aggregation and coagulation \cite{Fogelson:2008,BattistaIB2d:2017}. A fully parallelized IB method with adaptive mesh refinement was used, IBAMR \cite{BGriffithIBAMR}. More details on IB and IBAMR are found in the Appendix of \cite{Samson:coral-Re}.

\begin{figure}
\centering
\includegraphics[width=0.45\textwidth]{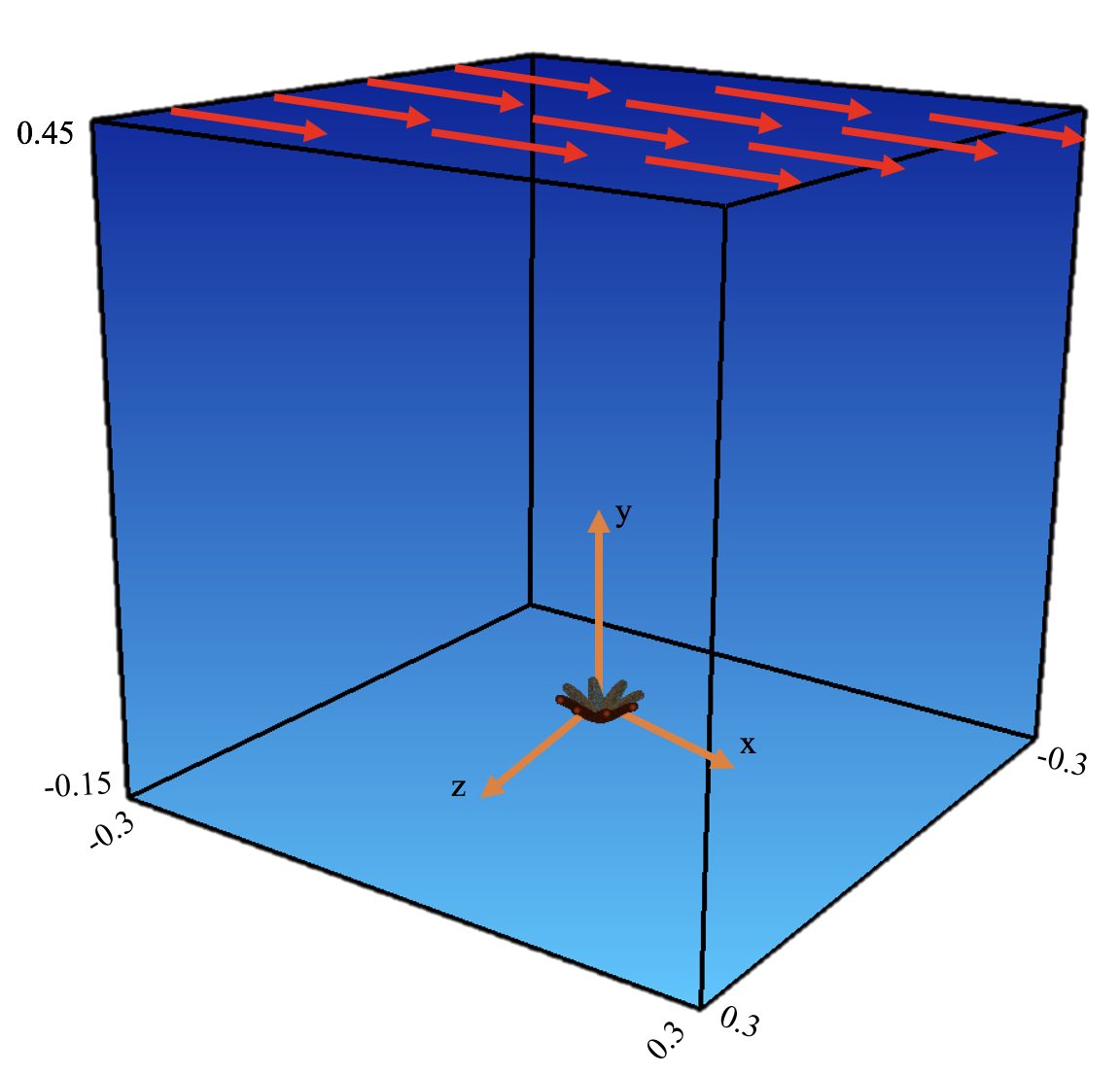}
\caption{The computational domain for a single coral polyp. Note that the $x$ and $z$ boundaries are periodic and the explicit boundary conditions on $y$ enforce an ambient shear flow.}
\label{fig:comp_domain}
\end{figure}

For the computational model, all physical and numerical parameters are given in Table \ref{table:num_param}. A depiction of the computational domain is shown in Figure \ref{fig:comp_domain}. The $x$ and $z$ boundaries have periodic boundary conditions while the $y$ boundaries are set up to induce an ambient shearing flow profile, $\textbf{u} = \textbf{0}$ at $y=-0.15$ and $\textbf{u} = [V_{max}\ 0\ 0]$ at $y=0.45$. For a study with no ambient flow, see \cite{Samson:coral-Re}. 

The Reynolds Number, $Re$, the ratio of inertial to viscous forces, is computed using the characteristic length of the tentacle, the characteristic pulsation frequency, and the density and dynamic viscosity of sea water. The biologically relevant $Re$ for a Xenia with measurements of $f_{coral} = 1/1.9\ s^{-1}$, $L_{coral} = 0.0045\ m$, $\rho=1023\ kg/m^3$ and $\mu= 0.00096\ kg/(ms),$ is 
\begin{equation}
    \label{Re_Xenia} Re = \frac{ \rho f_{coral} L^2_{coral} }{\mu} = 10.66.
\end{equation}

We considered the coral with background flows using a maximum velocity at the top of the fluid domain set to $V_{max}$ = 0, 0.05, 0.1, and 0.2 $dm/s$. Note that the velocity at the bottom of the domain was set to $V = 0\ m/s$. The shearing flow does not start off fully developed but ramps up during the first pulsation period of the coral, according to $V_{shear} = V_{max} \tanh\left(\frac{t}{T}\right).$

\begin{table}
\begin{center}
\begin{tabular}{| c | c | c | c |}
    \hline
    Parameter               & Variable    & Units        & Value \\ \hline
    Domain Size              & $D$        & m               &  $0.06$               \\ \hline
    Spatial Grid Size        & $dx$       & m               &  $D/1024$               \\ \hline
    Lagrangian Grid Size     & $ds$       & m               &  $D/2048$               \\ \hline
    Time Step Size           & $dt$       & s               &  $1.22\times 10^{-4}$   \\ \hline
    Total Simulation Time    & $T$        & \textit{pulses} &  $10$               \\ \hline
    Fluid Density            & $\rho$     & $kg/m^3$        &  $1000$               \\ \hline
    Fluid Dynamics Viscosity & $\mu$      & $kg/(ms)$       &  \textit{varied}      \\ \hline
    Ambient Flow Speed       & $V_{amb}$  & $m/s$           &  \textit{varied}      \\ \hline
    Tentacle Length          & $L_T$      & m               &  $0.0045$               \\ \hline
    Pulsing Period           & $P$        & s               &  $1.9$               \\ \hline
    Target Point Stiffness   &$k_{target}$&$kg\cdot m/s^2$  &  $9.0\times10^{-9}$  \\ \hline
    \hline
    \end{tabular}
    \caption{Numerical parameters used in the three-dimensional simulations.}
    \label{table:num_param}
    \end{center}
\end{table}

\begin{figure}
\centering
\includegraphics[width=0.5\textwidth]{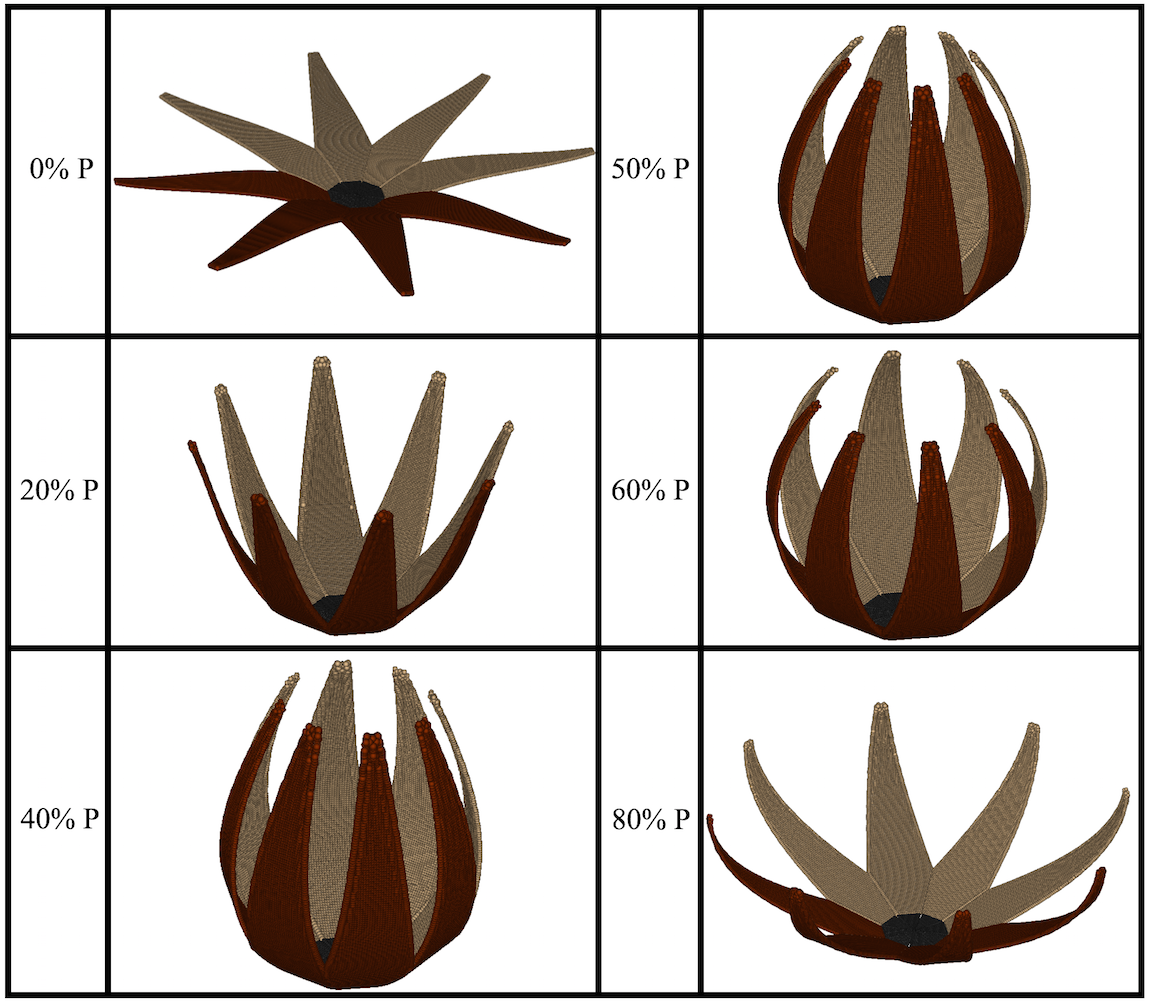}
\caption{A single polyp's pulsation cycle given by snapshots from different percentages of the pulsation period. The coral tentacles begins in a relaxed open state, then contract, and finally relax back to their open state. Note that as the polyp's tentacles relax, their tips appear more curved than compared to their contraction phase}
\label{fig:pulsation_cycle}
\end{figure}

To drive the coral's pulsing motion, the Lagrangian geometry of the tentacles was tethered to target points. The target points do not directly interact with the fluid and are moved in a prescribe fashion to best mimic the kinematics of the actual organism. The kinematics were captured by tracking positions along a single tentacle from 5 different coral polyps.
These positions were then fit with polynomials and averaged to enforce the prescribed motion of the immersed boundary, as in \cite{Samson:coral}. A coral pulsation cycle was divided into the 3 phases, see Figure \ref{fig:pulsation_cycle},  
\begin{enumerate}
    \item [1.] The coral begins with all of its tentacles in an open, relaxed state, and then actively contracts them until  reaching a closed state. This is the contraction phase.
    \item [2.] From the contracted state, the tentacles passively expand back to their original open state. This is the expansion phase.
    \item [3.] Once fully expanded, the tentacles may stay relaxed for some time until the next contraction. This is the relaxation phase.
\end{enumerate}

%
%

\section{Results}

\begin{figure}
\centering
\includegraphics[width=0.45\textwidth]{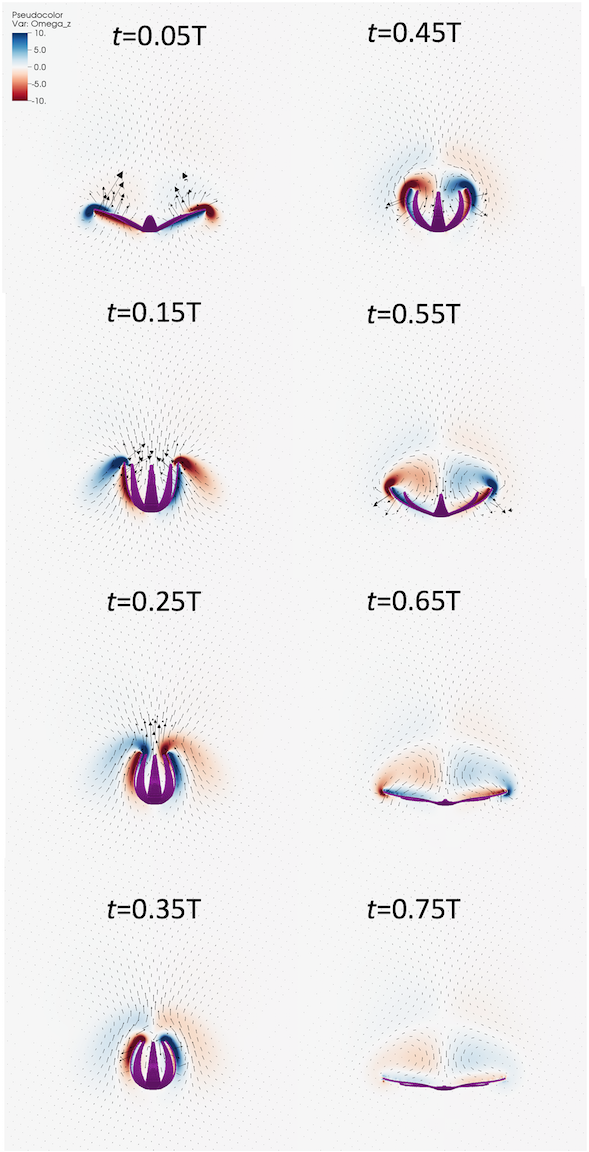}
\caption{The $z$-component of vorticity and the velocity vector field taken on a 2D plane through the central axis of the coral in a quiescent fluid. The colormap shows the value of $\omega_z$, the arrows point in the direction of flow, and the length of the vectors correspond to the magnitude of the flow. Snapshots are taken during the tenth pulse at times that are 5\%, 15\%, 25\%, 35\%, 45\%, 55\%, 65\%, and 75\% through the cycle.}
\label{fig:V0_snapshots}
\end{figure}

\begin{figure}
\centering
\includegraphics[width=0.45\textwidth]{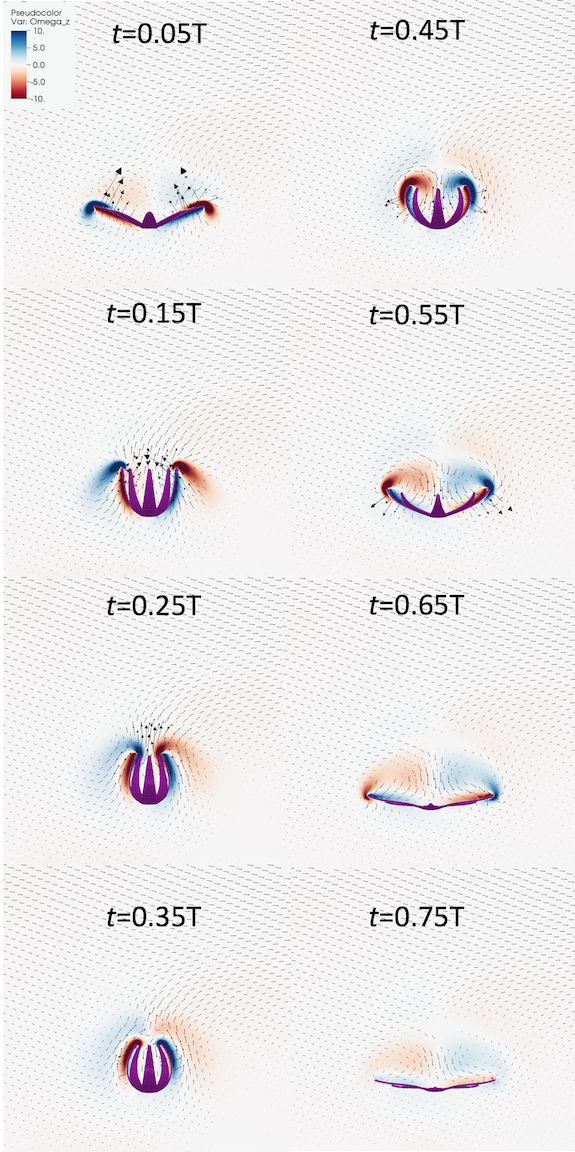}
\caption{The $z$-component of vorticity and the velocity vector field taken on a 2D plane through the central axis of the coral with shear background flow set to $V_{max}=0.05$ $dm/s$. The colormap shows the value of $\omega_z$, the arrows point in the direction of flow, and the length of the vectors correspond to the magnitude of the flow. Snapshots are taken during the tenth pulse at times that are 5\%, 15\%, 25\%, 35\%, 45\%, 55\%, 65\%, and 75\% through the cycle.}
\label{fig:V0p05_snapshots}
\end{figure}

\begin{figure}
\centering
\includegraphics[width=0.45\textwidth]{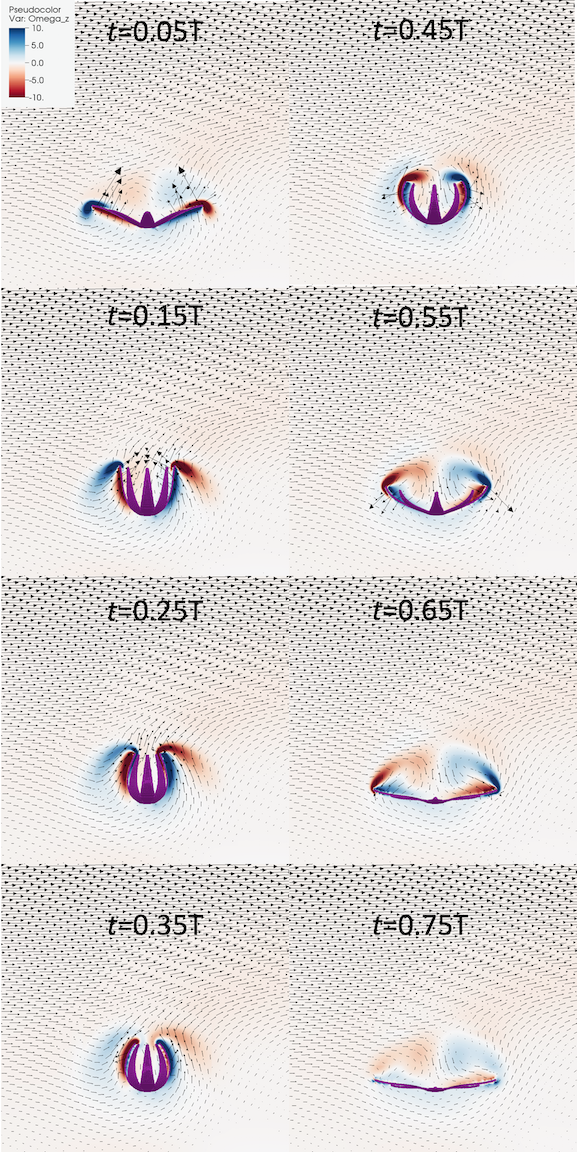}
\caption{The $z$-component of vorticity and the velocity vector field taken on a 2D plane through the central axis of the coral shear background flow set to $V_{max}=0.2$ $dm/s$. The colormap shows the value of $\omega_z$, the arrows point in the direction of flow, and the length of the vectors correspond to the magnitude of the flow. Snapshots are taken during the tenth pulse at times that are 5\%, 15\%, 25\%, 35\%, 45\%, 55\%, 65\%, and 75\% through the cycle.}
\label{fig:V0p2_snapshots}
\end{figure}

Figures \ref{fig:V0_snapshots}-\ref{fig:V0p2_snapshots} show snapshots of the velocity and vorticity generated during the tenth pulsation cycle for three different numerical simulations corresponding to ambient shear flows corresponding to $V_{max}= 0.0$, $0.05$, and $0.20$ $dm/s$, at the biologically relevant $Re = 10.66$. The velocity vectors point in the direction of flow, the length of the vectors correspond to the magnitude of the flow, and the colormap corresponds to the value of vorticity taken in the $z$-direction (out of plane). Both vorticity and fluid velocity were taken on a 2D plane passing through the central axis of the coral polyp. The tentacles are shown in pink in 3D. The snapshots taken correspond to 5\%, 15\%, 25\%, 35\%, 45\%, 55\%, 65\%, and 75\% of the pulsation period, such that the first three frames illustrate the contraction phase, the next four frames show the expansion phase, and the last frame shows the polyp at rest. Note that for roughly $35\%$ of a pulsation cycle the polyp is in its resting phase.

During contraction, regardless of ambient flow speed, there is an upwards jet produced. In addition, vorticity is generated at the tips of the tentacles during contraction. However, as the ambient flow increases, the upward jet begins to drift from completely vertical to off the $y$-axis in the direction of flow. Once contraction has stopped, the vortices formed at the tentacle tips separate and, depending on $V_{max}$, either quickly dissipate and the jet bends towards the direction of the ambient flow ($V_{max}=0.2\ dm/s$) or enhance the upwards jet ($V_{max}=0.0\ dm/s$).

When the polyp begins to relax and expand its tentacles ($t=0.35T$), oppositely spinning vortices form at the tips of each tentacle, compared to those formed during contraction. In the case of $V_{max}=0.0\ dm/s$, the vortices and flow fields remain symmetric throughout expansion; however, for $V_{max}>0\ dm/s$ the magnitude of vorticity is asymmetric. The vortex on the left side of the polyp is slightly larger than the vortex on the right, and the magnitude of the flow is also greater. Furthermore in the cases of $V_{max}>0\ dm/s$, the vortices are advected in the direction of the ambient flow. These vortices increase fluid mixing between the tentacles, in tandem with the background flow helping to contribute in bringing new fluid to within the polyp. 

While the polyp is at rest ($t=0.75T$), the weak vortices continue to pull fluid into the polyp. The flow is asymmetric for $V_{max}>0\ dm/s$, and symmetric for $V_{max}=0\ dm/s$. In the cases of non-zero ambient flows, the vortices are advected in the direction of the flow.

\begin{figure}
\centering
\includegraphics[width=0.475\textwidth]{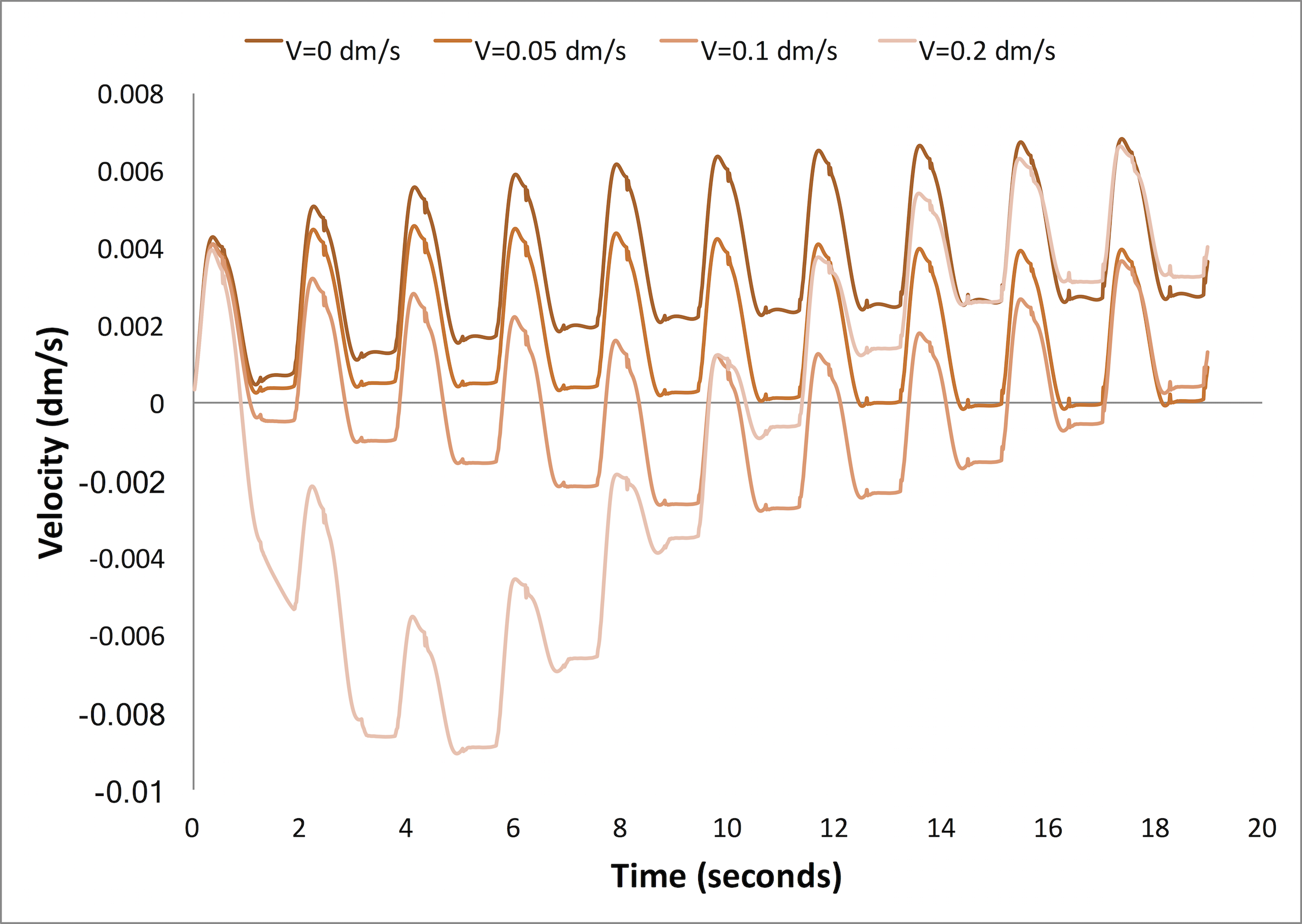}
\caption{The spatially averaged vertical flow above the polyp ($u_y$) versus time during the first ten pulse cycles. The maximal ambient shear flow was set to $V_{max}= 0$, $0.05$, $0.1$, and $0.2$ $dm/s$.}
\label{fig:vert_flow}
\end{figure}

Figure \ref{fig:vert_flow} shows the relative strength of the upward jet for several background velocities. The $y$-component of the velocity (in the vertical direction) was averaged within a box that was drawn from the tips of the tentacles during full contraction to one tentacle length above that point ($-0.0063 m <Y<-0.0018 m$). The width of the box was set equal to the diameter of the fully expanded polyp ($-0.0045 m<X,Z<0.0045 m$). The average vertical velocity versus time for ten pulses is provided for $V_{max}= 0$, $0.05$, $0.1$, and $0.2$ $dm/s$.

In each case of ambient flow considered we see a similar flow profile during the course of an individual pulsation cycle. During contraction upward flow velocity reaches a maximum, decreases during expansion, and remains constant while the polyp is in its resting state. However, we find that the ambient flow effectively suppresses the upward jet for $V_{max}>0\ dm/s$ until the background flow is fully developed. For example, for the strongest background flow considered, $V_{max}=0.2\ dm/s$, the flow profile continues to shift more negative until after the third pulsation cycle, when the profile begins to shift upward. The flow continues to become more positive until it almost surpasses the upward jet velocity corresponding to the case with no ambient flow by the tenth pulse. Upon comparing the flow profiles of the tenth pulse for $V_{max} = 0\ dm/s$ and $V_{max}=0.2\ dm/s$, we see that the upward jet during the expansion phase is greater in the case of no ambient flow. However, during the rest phase, the average upward jet velocity is greater in the case of $V_{max}=0.2\ dm/s$. Interestingly, the two other cases of ambient flow, $V_{max}= 0.05$ and $0.1$ $dm/s$, appear to take longer to reach periodic steady state than the maximum ambient flow. For $V_{max}=0.05\ dm/s$, the flow profile begins to shift upward after the sixth pulse, while for $V_{max}=0.1\ dm/s$, the profile begins to shift upward after the ninth pulse. These results suggest that for time varying ambient flows, the upward jets produced by the polyps would be highly dynamic.

\begin{figure}
\centering
\includegraphics[width=0.475\textwidth]{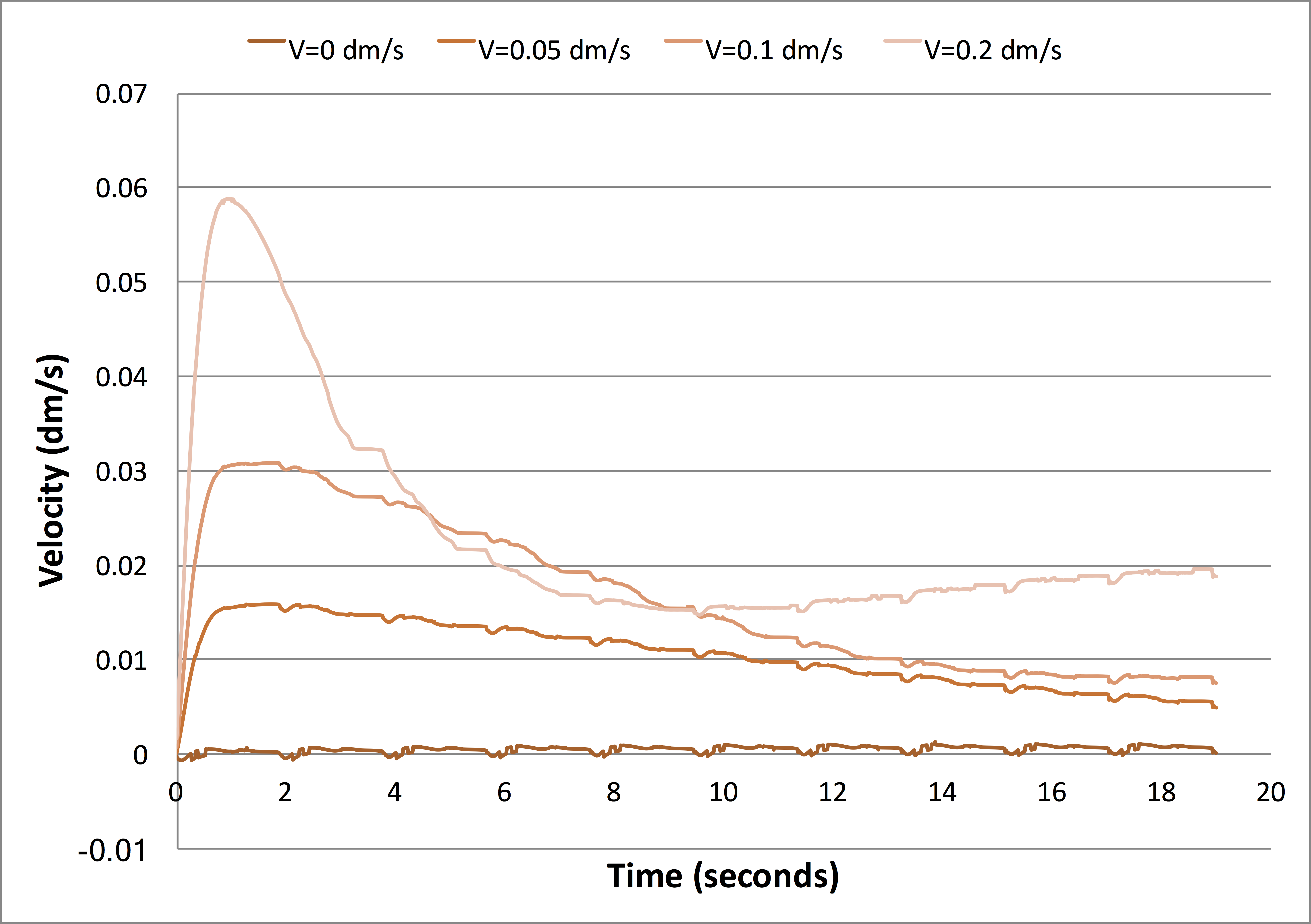}
\caption{The spatially averaged horizontal flow upstream of the polyp ($u_x$) over time during the first ten pulse cycles.  The maximum background flow was set to $V_{max}= 0$, $0.05$, $0.1$, and $0.2$ $dm/s$.}
\label{fig:hor_flow_left}
\end{figure}

\begin{figure}
\centering
\includegraphics[width=0.475\textwidth]{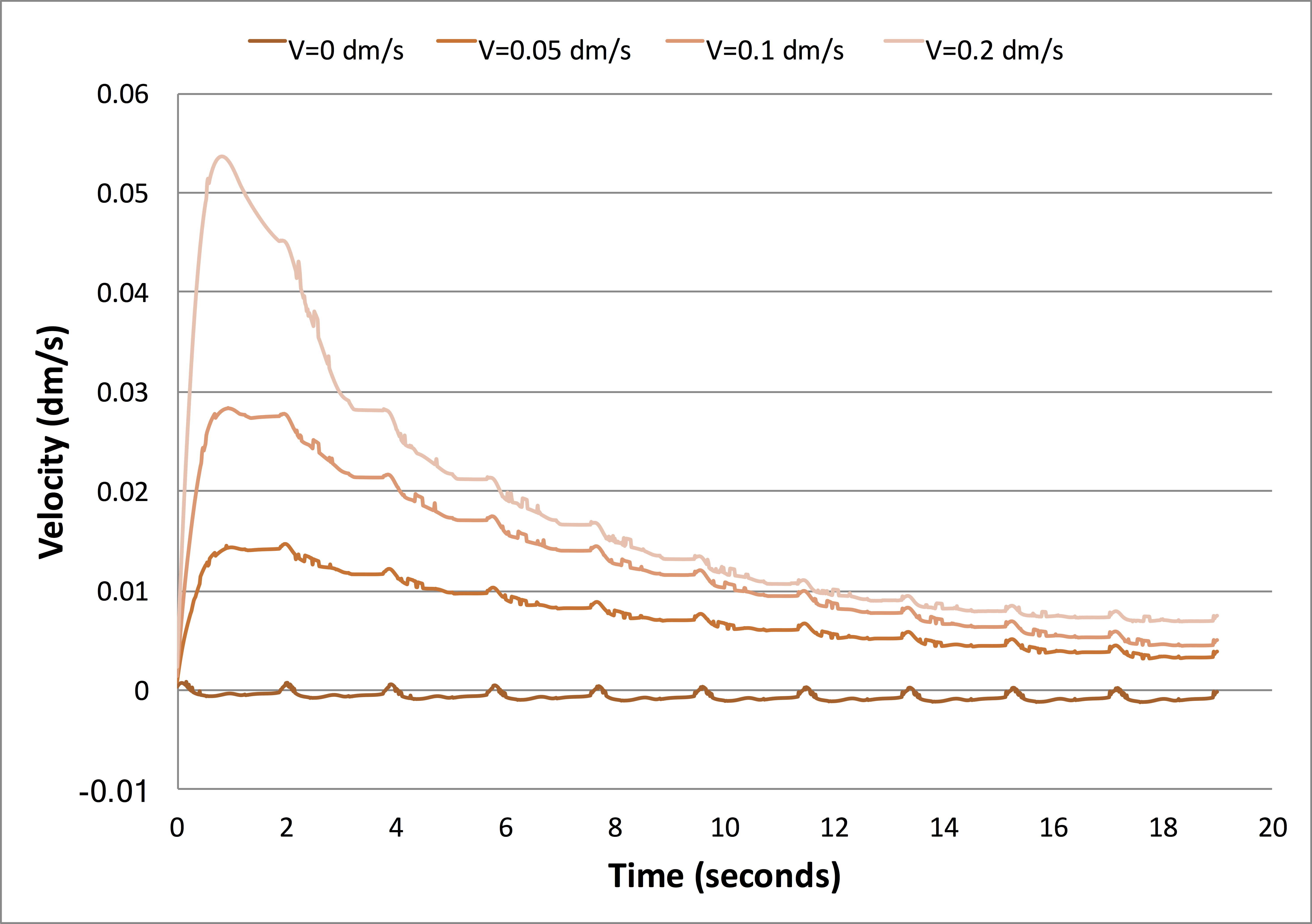}
\caption{The spatially averaged horizontal flow downstream of the polyp ($u_x$) over time during the first ten pulse cycles. $V_{max}= 0$, $0.05$, $0.1$, and $0.2$ $dm/s$.}
\label{fig:hor_flow_right}
\end{figure}

To compare the relative strength of the flow towards the polyp from the upstream and downstream directions, the $x$-component of the velocity was averaged within two boxes that were drawn from the tips of the tentacles during full expansion to one tentacle length to the left or right of that point ($-0.009 m<X<-0.0045 m$; $0.0045 m<X<0.009 m$). In the $z$-direction, the box was centered along the central axis of the polyp with a width about a ninth of its diameter ($-0.001 m<Z<0.001 m$). In the vertical direction, the box was drawn from the polyp base to the top of the fully contracted tentacle ($-0.01 m<Y<-0.0063 m$). The average horizontal velocity versus time upstream of the polyp is given in Figure \ref{fig:hor_flow_left}. Note that positive velocities are towards the polyp. Figure \ref{fig:hor_flow_right} shows the average horizontal velocity versus time downstream of the polyp for the ten pulses. Note that positive velocities are away from the polyp, and the background flow is in the positive direction.

Figure \ref{fig:hor_flow_left} illustrates that in the case of no ambient flow there is relatively little flow in the direction of the polyp. When there is background flow, e.g., $V_{max}>0\ dm/s$, there is significant flow in the direction of the polyp during the first ten pulses. In all cases during the first pulsation cycle, the average velocity reaches a maximum around the time that the polyp finishes its first contraction. Note that during this time the ambient flow is developing from rest at $t=0$. After the first contraction, the flow velocity toward the polyp decreases. This is due to the added drag due to the presence of the coral and its wake. For $V_{max}=0.2\ dm/s$, the average velocity reaches a minimum after the fifth pulse and then begins to increase once again. It is possible that this trend would also occur for the other two non-zero ambient flow cases since periodic steady state has not been reached after 10 pulses. Given the long times to reach steady state, these results suggest that unsteady flows, similar to the flow fields these corals would experience in their natural habitat, would introduce complex transient fluid behavior.  

Figure \ref{fig:hor_flow_right} shows overall similar trends as in Figure \ref{fig:hor_flow_left}, but with a few subtle differences. While there is minimal horizontal flow downstream of the polyp for the case of no ambient flow, the motion of the tentacles does pull fluid towards the polyp (negative direction) during most of the pulsing cycle. For $V_{max}>0\ dm/s$, as the ambient flow develops during the first pulse, the average velocity reaches a maximum approximately when the polyp finishes the contraction phase. In comparing these flow speeds to those in Figure \ref{fig:hor_flow_left}, the velocities are slightly smaller. Moreover, the horizontal flow decreases after reaching its maximum and continue to decrease during the 10 pulses. Note that for the ten pulsation cycles of data, in no case did the average velocities begin to increase again, such as in the $V_{max}=0.2\ dm/s$ in Figure \ref{fig:hor_flow_left}. These differences may be attributed to the vortices produced by the polyp during contraction and expansion that are advected with the flow.

%
%

\section{Conclusion}

In this paper, we use immersed boundary simulations to reveal the interactions between the flow currents generated by pulsing soft corals and a background shear flow. We find complex interactions between these two flows. The addition of a background current may drive fluid into the polyp or enhance the upward jet above the polyp depending on the strength of the flow. In general, the pulsing motion tends to reduce the magnitude of the horizontal flow near the polyp over time. Similar to the upside down jellyfish \cite{Hamlet:2014}, this reduced horizontal flow may allow for slow sampling of the fluid and enhanced particle uptake. These flows also take a long time to fully develop ($>10$ pulse cycles), suggesting that transient dynamics are important to the dynamics of the jet. To fully describe the interaction between passive and active exchange currents, additional numerical, analytical, and experimental studies are warranted that consider multiple coral polyps in complex, time varying flows.

%
%

\section*{Acknowledgment}

The authors would like to thank Uri Shavit and Roi Holzman for introducing us to pulsing soft corals and for their assistance in the field and the organizers of the 2017 BIOMATH meeting at Kruger Park, South Africa. The authors would also like to acknowledge funding from NSF PHY grant \#1505061 (to S.K.) and \#1504777 (to L.A.M.), NSF DMS grant \#1151478 (to L.A.M.), and NSF DMS grant \#1127914 (to the Statistical and Applied Mathematical Sciences Institute). Travel support for J.E.S. was obtained from the Company of Biologists, and J.E.S. was supported by an HHMI International Student Research Fellowship and the Women Diver's Hall of Fame. 

%
%

\bibliographystyle{elsarticle-num}
\bibliography{coral}

\end{document}